%% file: lrjohnson_bact_aging2005.tex
\documentclass[12pt]{article}
\usepackage{fullpage}
\usepackage{spec}
\usepackage{epsfig}
\usepackage{epsf}
\usepackage{amsmath}
\usepackage[footnotesize]{caption}
\usepackage{setspace} 
\include{lamac}

\begin{document}
\title{Implications of Aging in Bacteria}
\author{
Leah R. Johnson$^{1*}$ and Marc Mangel$^2$\\
{\small $^1$ Department of Physics}\\
{\small $^2$ Department of Applied Mathematics and Statistics, }\\ 
{\small University of California, Santa Cruz, 1156 High St., Santa Cruz, CA 95064}\\
{\small$^*$ Corresponding author. phone: 831-459-5385;  E-mail: {\tt lrjohnson@ams.ucsc.edu}}
\date{}
}
\maketitle


\pagebreak

\begin{abstract}
Motivated by recent research of aging in {\it E. coli}, we explore the effects of aging on bacterial fitness. The disposable soma theory of aging was developed to explain how differences in lifespans and aging rates could be linked to life history trade-offs. Although generally applied for multicellular organisms, it is also useful for exploring life history strategies of single celled organisms such as bacteria. Starting from the Euler-Lotka equation, we propose a mathematical model to explore how a finite lifespan effects fitness of bacteria. We find that that there is surprisingly little loss of fitness when the bacterium has limited opportunities to reproduce. Instead, the fitness gained each time the bacteria reproduces decreases rapidly, and investing resources to survive to reproduce the first few times is likely more advantageous than investing additional resources to try to maintain cell integrity longer.
\end{abstract}

\pagebreak
 
\section{Introduction}

Living organisms exhibit an astonishing variety of lifespans and life histories. A giant sequoia tree can live for thousands of years, producing only a few offspring over its lifetime, while a dandilion survives only for a season, but produces thousands of seeds, often resulting in hundreds of new plants the next year. Among a single genus, such as rockfish ({\it Sebastes}), species exhibit lifespans ranging  anywhere from 10 to 200 years \cite{love}. 
Even bacteria appear to age. New research by Stewart et al.~\cite{stewart:2005} indicates that even {\it Escherichia coli} which appear to divide symmetrically actually divide into one ``old'' and one ``young'' cell. 
Current theories of aging seek to combine principles of evolution with theories from microbiology about how damage that accumulates in a cell makes it unable to function. One such theory is called the disposable soma theory of aging \cite{kirkwood1,finch,drenos:2005}. This model predicts that 
an organism will allocate resources for reproduction and maintainace of the soma 
in such a way that maximizes the dispersal of its genes. We expect that the optimal allocation strategy will not be one that allows an organism to maintain the soma indefinitely. This is because organisms have a finite amount of energy to use for all life functions; if they use all of the energy for maintenance of the soma, then there might not be enough energy to reproduce, and vice versa. 
This theory also predicts how natural mortality is expected to influence the life span and reproductive schedule of an organism. For instance, if an organism experiences high natural mortality, then its resources are better invested in offspring than soma. Higher natural mortality also results earlier maturation, so they will be less likely to die before having an opportunity to reproduce. Organisms with lower natural mortality instead maintain the soma for longer, produce offspring less frequently, and experience longer lives. 

\section{Fitness Models of Bacteria} \label{section2}
Single celled organisms, or other organisms where the soma and the germ line are not separate, were once expected to be immortal \cite{williams:1957}. The distinction between organisms that age and those that do not has more recently been hypothesized to depend upon asymmetry in reproduction \cite{partridge:1993}. 
Recent research by Stewart et al.~\cite{stewart:2005} indicates that even bacteria that appear to divide symmetrically, such as {\it E. Coli}, actually produce functionally asymmetric cells during cell division. They consider one of the cells to be an aging parent cell that produces offspring that are ``rejuvenated'' and find evidence evidence that these older cells reproduce more and more slowly as they age, and may even stop reproducing after a certain number of divisions \cite{stewart:2005}.  Quantitative models can be useful for exploring how evolutionary trade-offs shape aging and senescence in these simple organisms.  Here we explore the effect of finite lifespan on bacterial fitness when doubling rate and environment are constant. 

We explore the effects of life history choices on our selected fitness measure, the intrinsic rate of natural increase, denoted by $r$, in populations with age dependent reproduction and mortality schedules, living in a constant environment using the Euler-Lotka equation:
\begin{align}
1 &= \int_0^{\infty} e^{-rx} l_x b_x dx. \label{eulerlotka}
\end{align}
Here the probability of surviving to age $x$ is denoted as $l_x$ and the number of  
offspring born to an individual
of age $x$ is $b_x$.

\begin{figure}[h!]
\begin{center}
\includegraphics[scale=0.2]{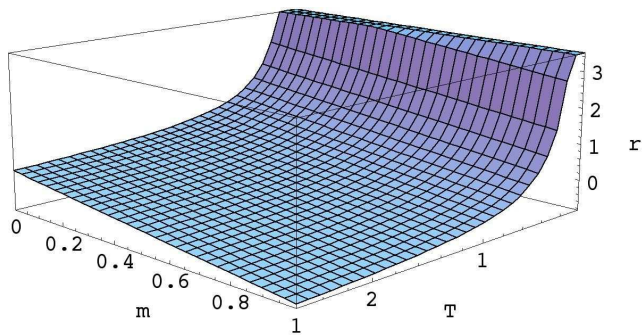}  \\
\end{center}
\caption{The intrinsic rate of natural increase, $r$,  as a function of the doubling time, $T$, and the mortality, $m$. \label{plot:r_vs_MandT}}
\end{figure}

Kirkwood \cite{kirkwood1} first proposed a simple model of bacterial fitness for cells that divide perfectly symmetrically, based upon (\ref{eulerlotka}), with appropriate choices of $b_x$ and $l_x$ for a clonally reproducing population.
One simple choice for $l_x$ is an exponentially decreasing survival probability, $l_x = e^{-mx}$, where $m$ is the mortality rate. The birth rate will be related to the doubling time, $T$. If an individual bacteria survives to time $T$, it divides, and two identical offspring result. At the end of this time period the original bacteria is essentially ``dead'' and the daughter cells remain. An appropriate ``birth'' rate would therefore be $b_{x} = 2 \delta(x-T)$, where $\delta(x-T)$ is the Dirac delta function\footnote{The Dirac delta function is defined as a unit impulse at some point $x_0$ such that: 
\begin{align*}
\delta(x-x_0) = 0, \hspace{0.5cm} & x \neq x_0\\
\int_{-\infty}^{\infty} \delta(x-x_0) dx & = 1,\\
\intertext{and given an arbitrary function $f(x)$:}
\int_{-\infty}^{\infty} f(x) \delta(x-x_0) dx & = f(x_0).
\end{align*}
}. 
With these expressions for $l_x$ and $b_x$ (\ref{eulerlotka}) becomes:
\begin{equation}\label{EL2}
2e^{-(m+r)T}=1.
\end{equation}
Figure \ref{plot:r_vs_MandT} shows this functional relationship between $r, m$ and $T$ graphically. As $m$ and $T$ increase, $r$ decreases. As $T \rightarrow 0 , r \rightarrow \infty $, regardless of the value of $m$. As $m$ and $T$ increase, $r$ decreases.  

We denote the fraction of resources allocated for growth and reproduction by $\rho$, and the fraction alloted for maintenance and survival by $1-\rho$. Following Kirkwood \cite{kirkwood1}, we can parameterize the mortality, $m$, and doubling time, $T$, in terms of $\rho$ as:
\begin{eqnarray} 
T(\rho) &=&\frac{ T_{0}}{\rho} \label{eq:T}\\
m(\rho) &=& \frac{m_0}{1-\rho} \label{eq:m}.
\end{eqnarray} 
Here, $T_{0}$ can be thought of as the minimum possible time it would take for the bacteria to reproduce if it allocated all of it's resources to growth; $m_0$ is the minimum mortality of the bacteria if all resources are allocated for survival. 
Solving for $r$ in (\ref{EL2}) with the expressions for $T$ and $m$ in (\ref{eq:T}-\ref{eq:m}) yields:
\begin{equation}
r= \frac{\rho}{T_0} \ln2 - \frac{m_0}{1-\rho}. \label{eq:r_Tm}
\end{equation}
Maximizing (\ref{eq:r_Tm}) with respect to $\rho$ gives the optimal resource allocation, $\rho^*$:
\begin{equation}
\rho^* = 1-\left( \frac{T_0 m_0}{\ln2} \right)^{\frac{1}{2}}, \label{rhostar}
\end{equation}
and the corresponding value of $r_{max}$,
\begin{equation}
r_{max} = \frac{1}{T_0} \left( \ln{2} - 2(m_0 T_0 \ln{2})^{\frac{1}{2}} \right). \label{rmax} 
\end{equation}

\begin{figure}[h!]
\begin{center}
\begin{tabular}{cc}
\includegraphics[scale=0.225]{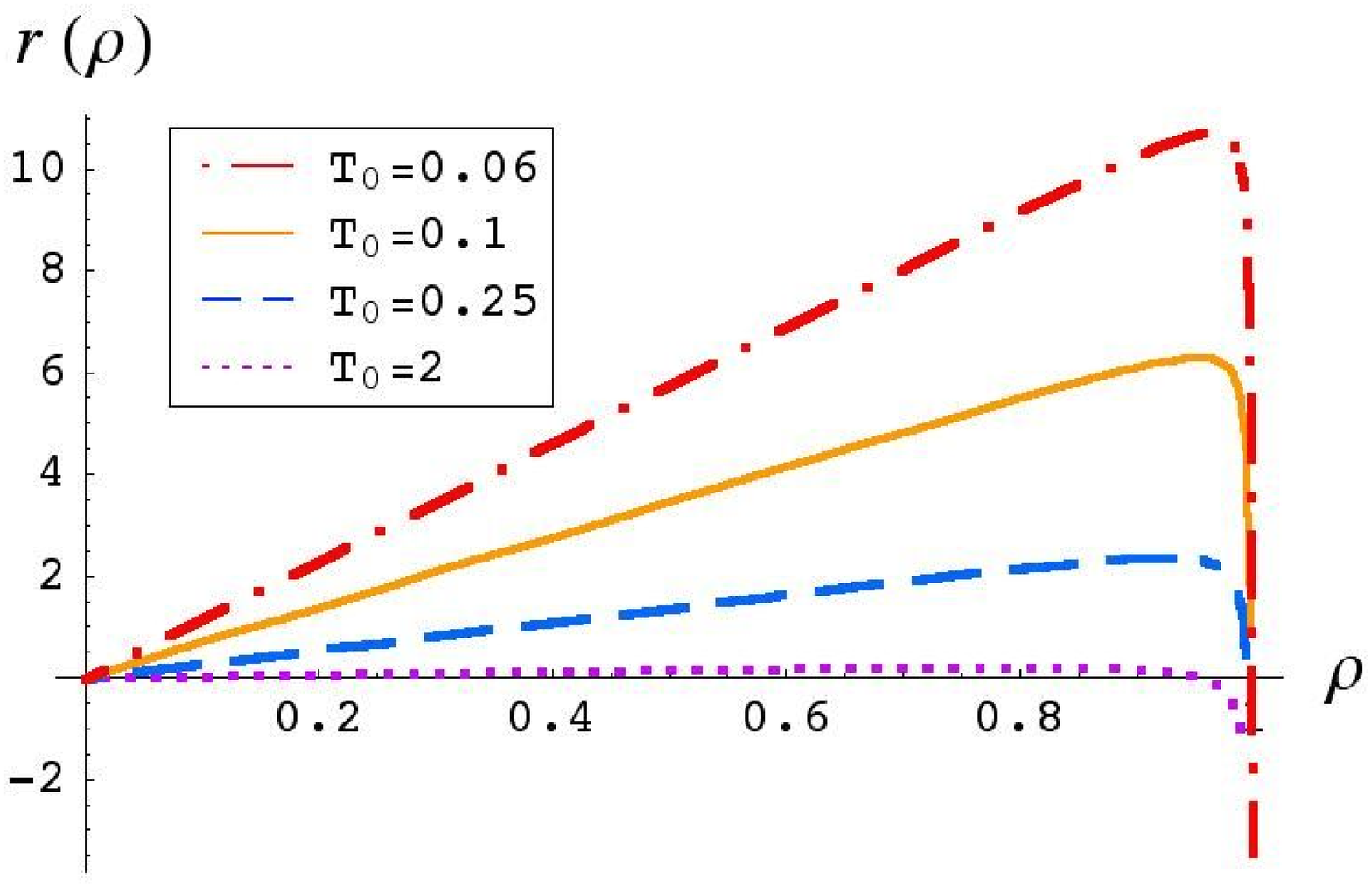}  \\
(a) \\
\includegraphics[scale=0.225]{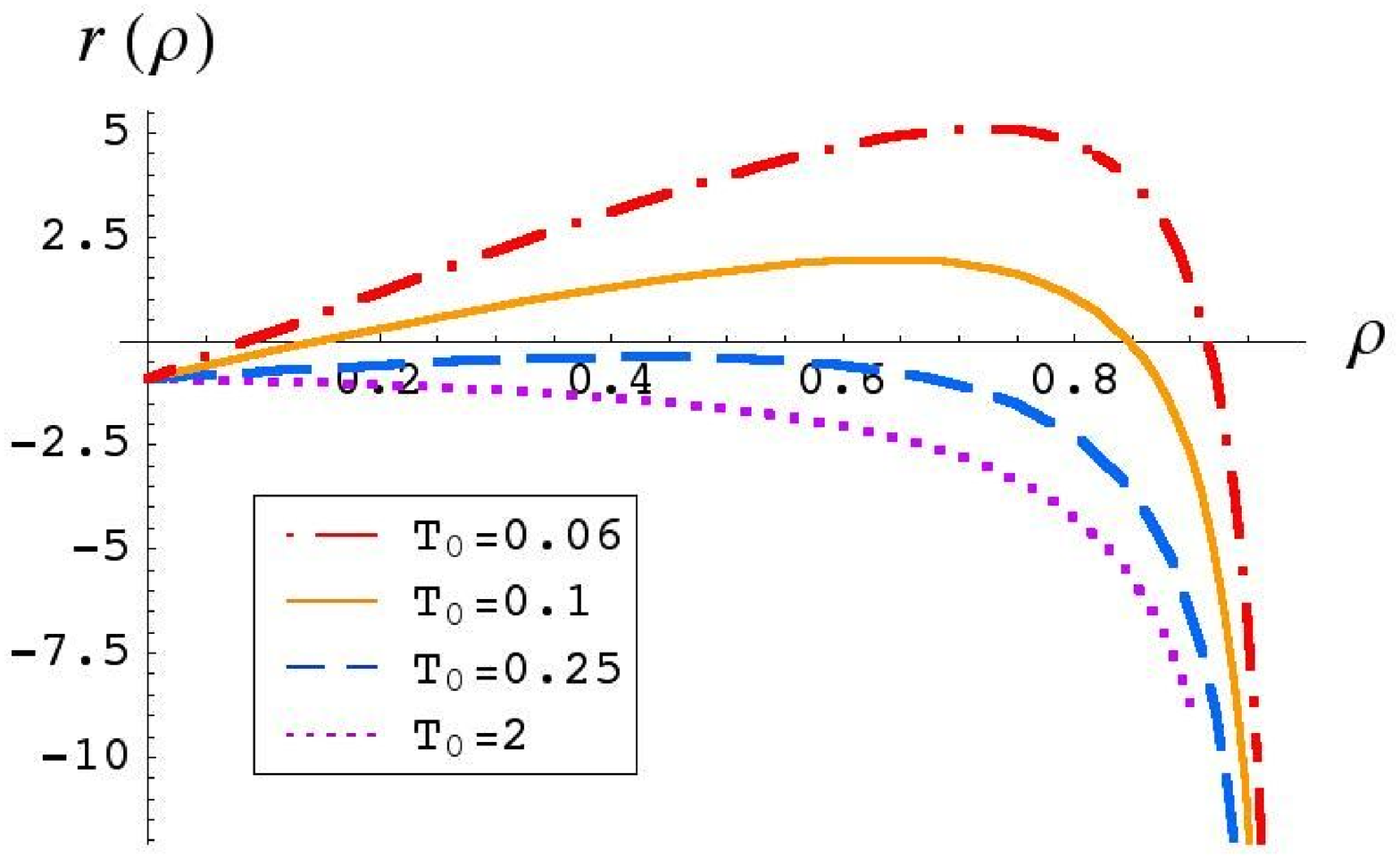} \\
(b) \\
\end{tabular}
\end{center}
\caption{$r$ as a function of $\rho$ for four values of $T_0 = (2, 0.25, 0.1, 0.06)$ for (a) $m_0=0.015$ (b)  $m_0=0.9$. High values of $T_0$ correspond to the lowest curves in each plot, and small $T_0$ to higher curves. \label{plot:r_vs_rho}}
\end{figure}

If $T_0$ and $m_0$ are constants, $r(\rho)$ will follow curves similar to those depicted in Figure \ref{plot:r_vs_rho}. We can see from Eqn. (\ref{rhostar}) and Figure \ref{plot:r_vs_rho} that the optimal strategy will never be to dedicate all resources to reproduction, i.e. $\rho^* < 1$. In an environment with low mortality, it can be optimal to invest most resources in reproduction, as shown in Figure \ref{plot:r_vs_rho}a, even if the generation time is relatively long (lowest curve). However, if the mortality is high and/or the doubling time is long, the optimal resource allocation may be to use more resources for survival (Figure \ref{plot:r_vs_rho}b).

\section{The Implications of Bacterial Aging}
We now explore how the intrinsic rate of natural increase for the aging bacteria, denoted by $\tilde{r}$, is impacted by a finite lifespan.
Starting from the Euler-Lotka Eqn. (\ref{eulerlotka}) we first modify the birth rate, $b_x$, to take into account a functional asymmetry in cell division. The cell is able to divide and produce a single offspring in a given fixed doubling time $T$, then can live to divide again later. We define cellular ``age'', $a = 1, 2, \dots , a_{max}$, as the number of times the cell has doubled, where $a_{max}$ is the maximum number of times it can split (in analogy with the Hayflick limit \cite{arking,finch}). The birth rate is then a sum of delta functions spaced a distance $T$ apart:
\begin{equation} \label{eq:BR1}
b_x = \sum_{a=1}^{a_{max}} \delta{(x-aT)}.
\end{equation} 
This, together with the previous expression for $l_x$, inserted into (\ref{eulerlotka}) gives:
\begin{align}
1 & = \int_0^{\infty} e^{-\tilde{r}x} e^{-mx} \sum_{a=1}^{a_{max}} \delta{(x-aT)} dx \nonumber\\
&= \sum_{a=1}^{a_{max}} e^{-(\tilde{r}+m)aT}. \label{eq:EL3}
\end{align}
 Evaluating the sum in Eqn. (\ref{eq:EL3}) results in
\begin{equation}
e^{(\tilde{r}+m)T} = 2- e^{-(\tilde{r}+m)Ta_{max}}. \label{eq:EL_final1}
\end{equation}

\begin{figure}[h!]
\begin{center}
\begin{tabular}{c}
\includegraphics[scale=0.275]{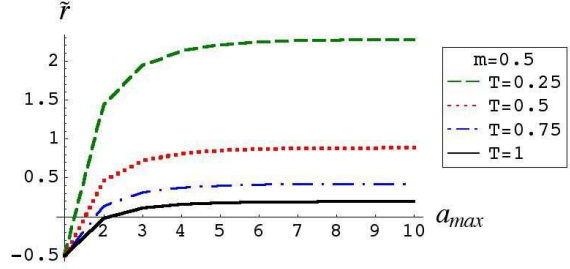}\\
(a) \\
\includegraphics[scale=0.275]{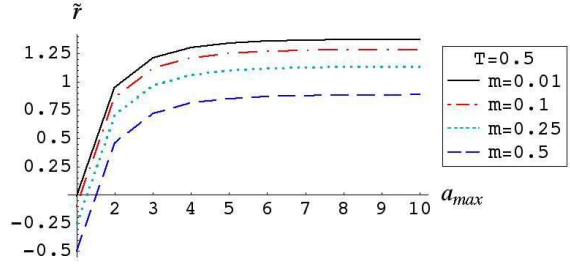} \\
(b) \\
\end{tabular}
\end{center}
\caption{$\tilde{r}$ vs $a_{max}$ for (a) $m=0.5$, $T=(0.25, 0.5, 0.75, 1)$ and (b) $T=0.5$ and $m=(0.5, 0.25, 0.1, 0.01)$ (c) Effects of $a_{max} = (2, 3, 5, 10)$ on $\tilde{r}(T)$ for $m= 0.9$} \label{fig:tilde_r_vs_amax} 
\end{figure}

This equation has two solutions for $a_{max} \ge1$. The trivial solution exists when $\tilde{r} = -m$. All other solutions depend upon the maximum age, $a_{max}$, the mortality rate, $m$, and the doubling time, $T$ (see Figure \ref{fig:tilde_r_vs_amax}). The value of $\tilde{r}$ varies considerably depending upon the combination of these three parameters.  Variation of $T$ and $a_{max}$ seem to have the most impact upon $\tilde{r}$, as shown in Figure \ref{fig:tilde_r_vs_amax}a, whereas variations in $m$ act to shift $\tilde{r}$ up or down a fixed amount when $T$ is held constant (Figure \ref{fig:tilde_r_vs_amax}b). Variation in $a_{max}$ has a much larger impact upon $r$ when the value of $a_{max}$ is small. When $a_{max} \rightarrow \infty$,  Eqn. (\ref{eq:EL_final1}) reduces to Eqn. (\ref{EL2}), and $\tilde{r} \rightarrow r$. In other words, infinite reproductive lifespan and perfectly symmetrical cell division are equivalent. 

Since as $a_{max}$ increases $\tilde{r} \rightarrow r$, we can approximate the second solution of Eqn. (\ref{eq:EL_final1}) in terms of $r$ for large values of $a_{max}$ using Newton's method:
\begin{equation}\label{eq_Rapprox}
\tilde{r} \approx r - \frac{1}{T} \frac{2^{-a_{max}}}{\ln{2} + a_{max} 2^{-a_{max}}}.
\end{equation}

\begin{figure}[h!]
\begin{center}
\begin{tabular}{c}
\includegraphics[scale=0.275]{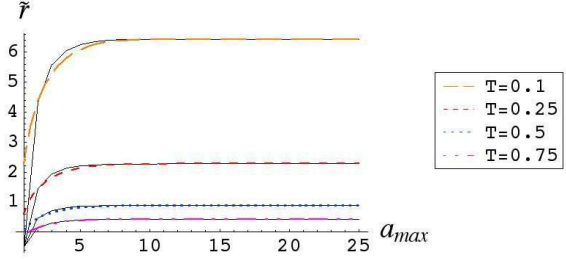}  \\
 (a) \\
\includegraphics[scale=0.275]{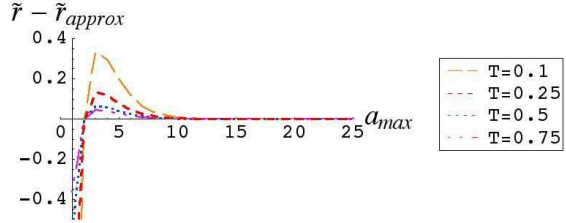} \\
(b)  \\
\end{tabular}
\end{center}
\caption{(a) Analytic approximation (dashed lines) and numerical solution (solid lines) of $\tilde{r}$ vs $a_{max}$ for $m=0.5$ and $T= (0.1, 0.25, 0.5, 0.75)$  (b) The difference between the numerical solution and analytic approximation, $\tilde{r} - \tilde{r}_{approx}$, for the same values as in (a) \label{fig:r_tilde_approx}}
\end{figure}

As can be seen in Figure \ref{fig:r_tilde_approx}, this approximation is remarkably good, over a wide range of values of $a_{max}$. It also gives us some idea of the values of $a_{max}$ that are ``large''. Where the approximation is valid, $a_{max}$ must be large enough to make the difference in fitness between the aging bacteria and the immortal bacteria, $r-\tilde{r}$, very small.

Since the fitness of the aging and immortal bacteria are very close when $a_{max}$ is large, we would expect that the allocation strategies would be similar as well. Let us denote the allocation strategy of an aging bacteria by $\tilde{\rho}$. Substituting (\ref{eq:T}) and (\ref{eq:m}) into (\ref{eq:EL_final1}) gives a relationship between $\tilde{r}$, $\tilde{\rho}$, $m_0$, $T_0$, and $a_{max}$:
\begin{equation}
\exp{\left((\tilde{r}+\frac{m_0}{1-\tilde{\rho}})\frac{T_0}{\tilde{\rho}}\right)} = 2- \exp{\left(-(\tilde{r}+\frac{m_0}{1-\tilde{\rho}})\frac{T_0}{\tilde{\rho}}a_{max}\right)}. \label{eq:EL_allocate1}
\end{equation}

\begin{figure}[h!]
\begin{center}
\begin{tabular}{ccc}
\includegraphics[scale=0.2]{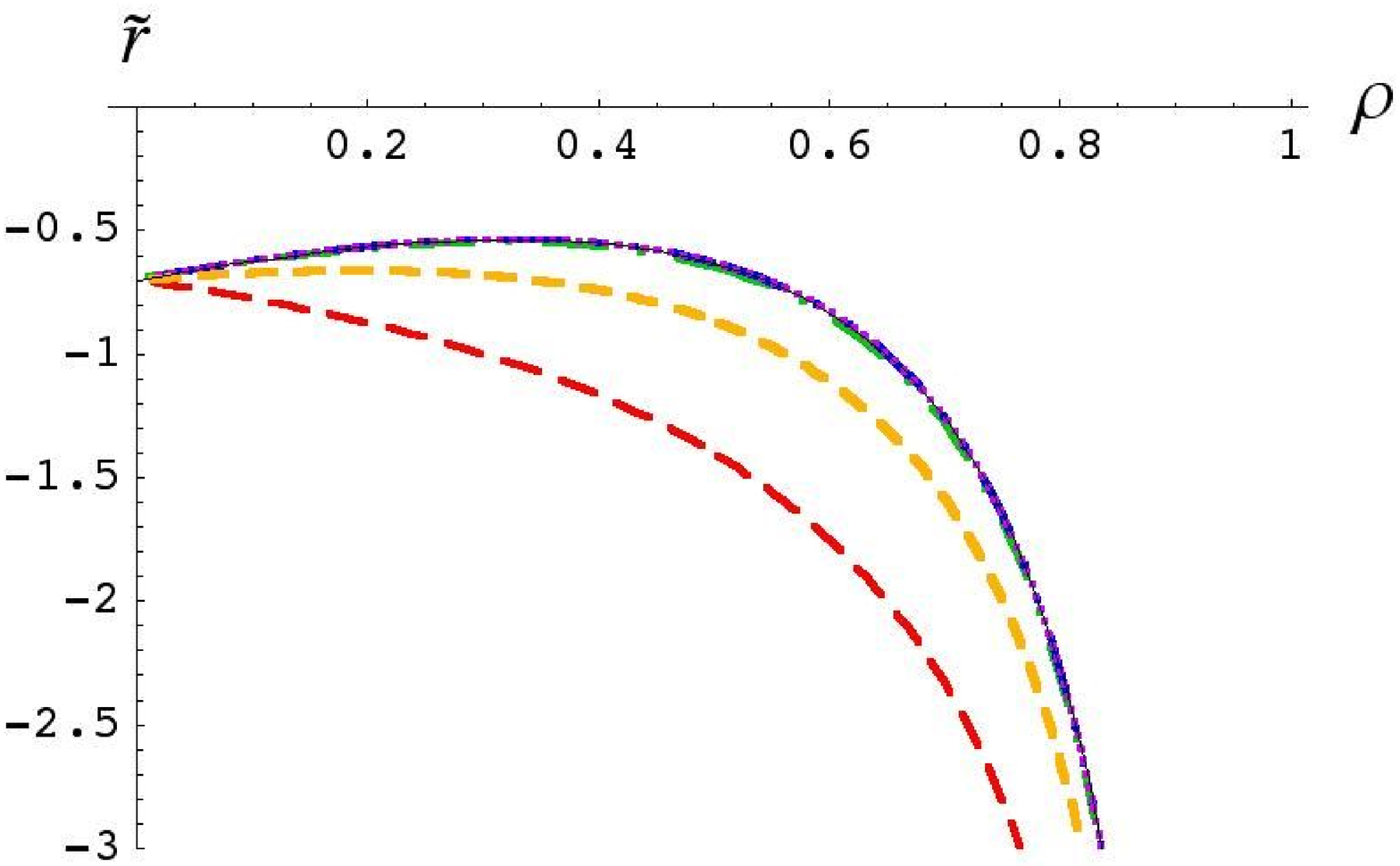}&
\includegraphics[scale=0.2]{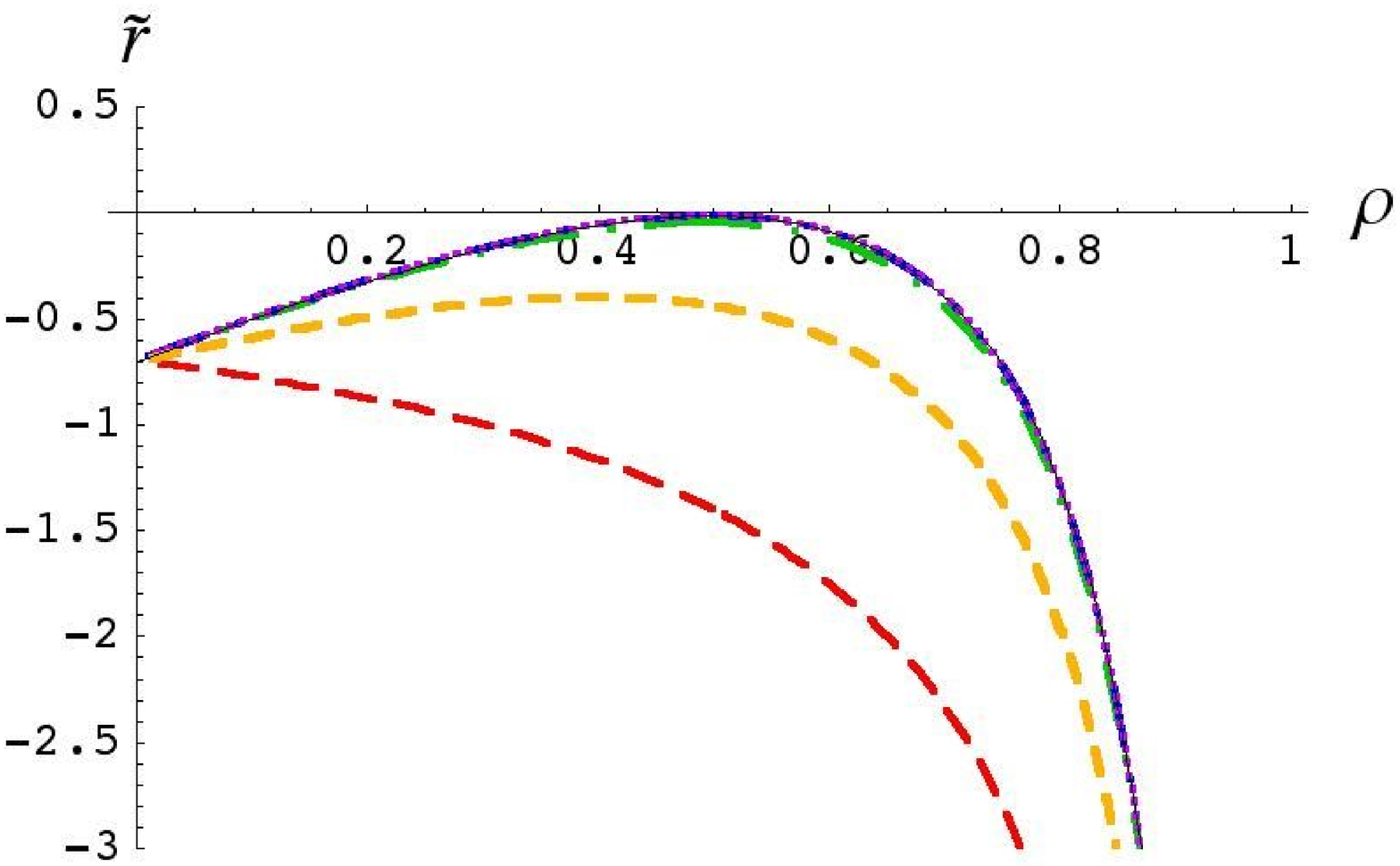}& \\
(a) & (b) & \\
\includegraphics[scale=0.2]{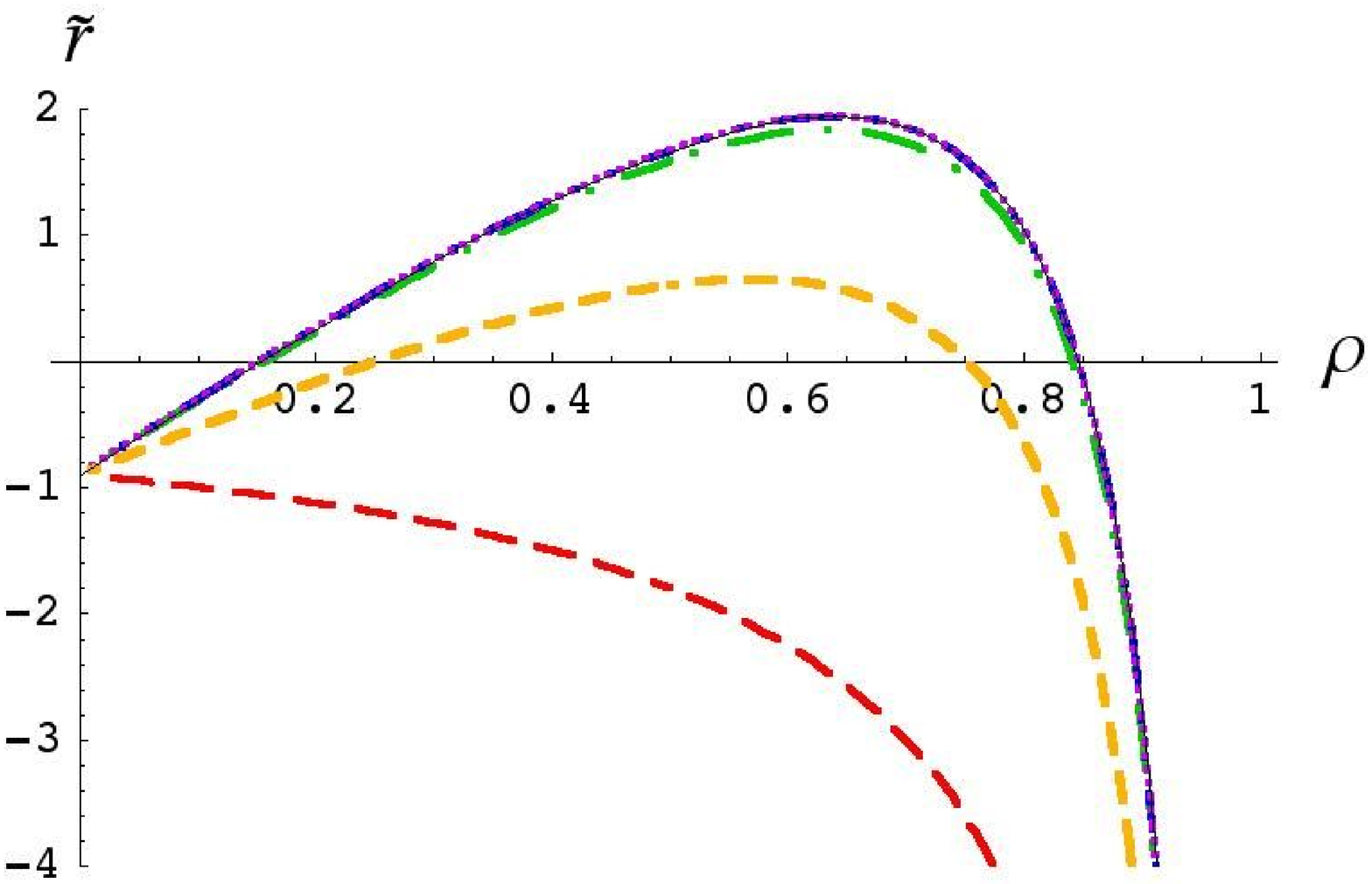}&
\includegraphics[scale=0.2]{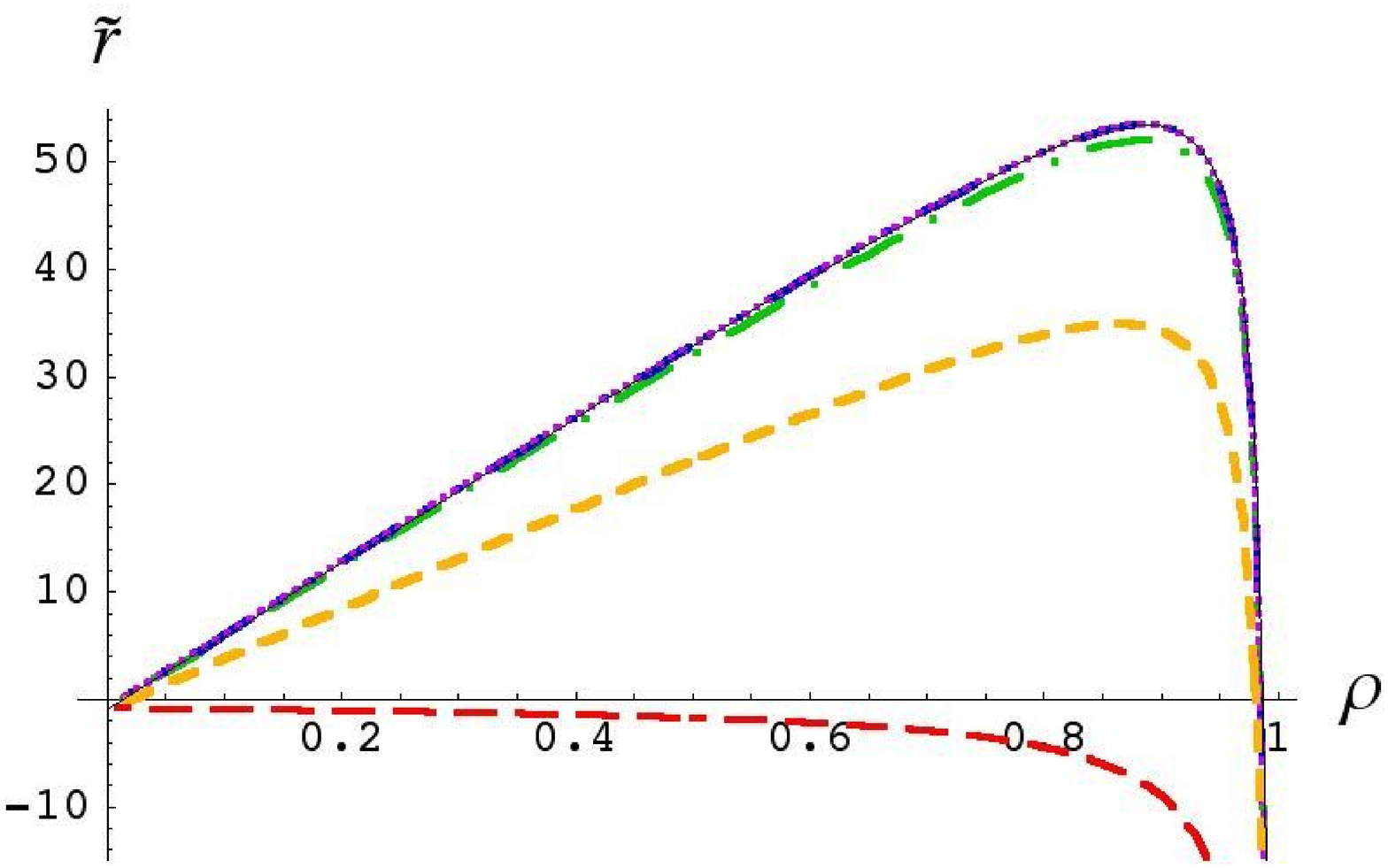}&
\includegraphics[scale=0.35]{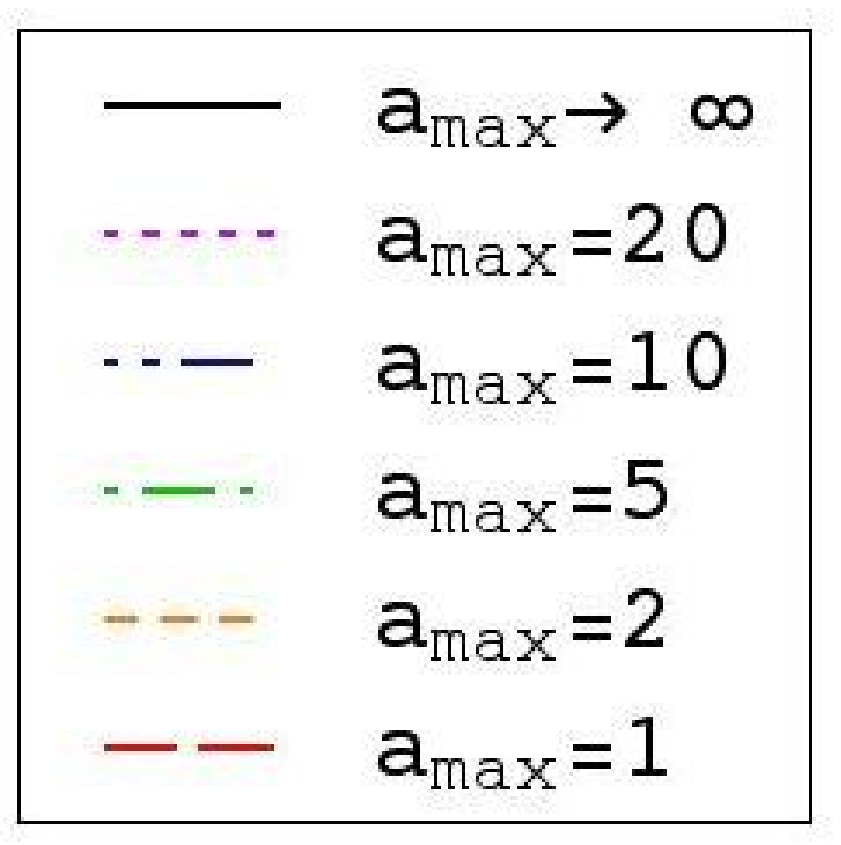}\\
(c) & (d)  &\\
\includegraphics[scale=0.2]{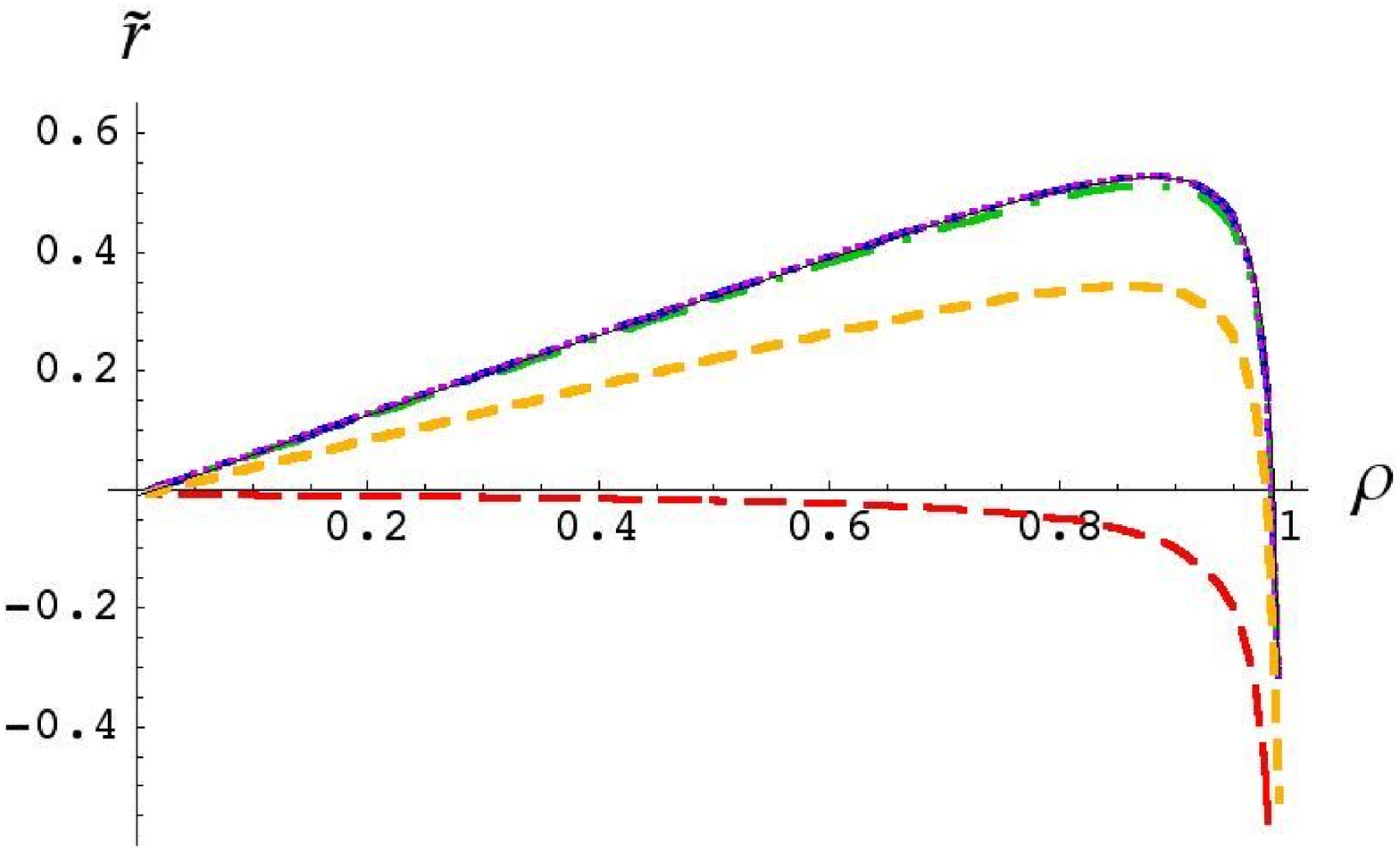}&
\includegraphics[scale=0.2]{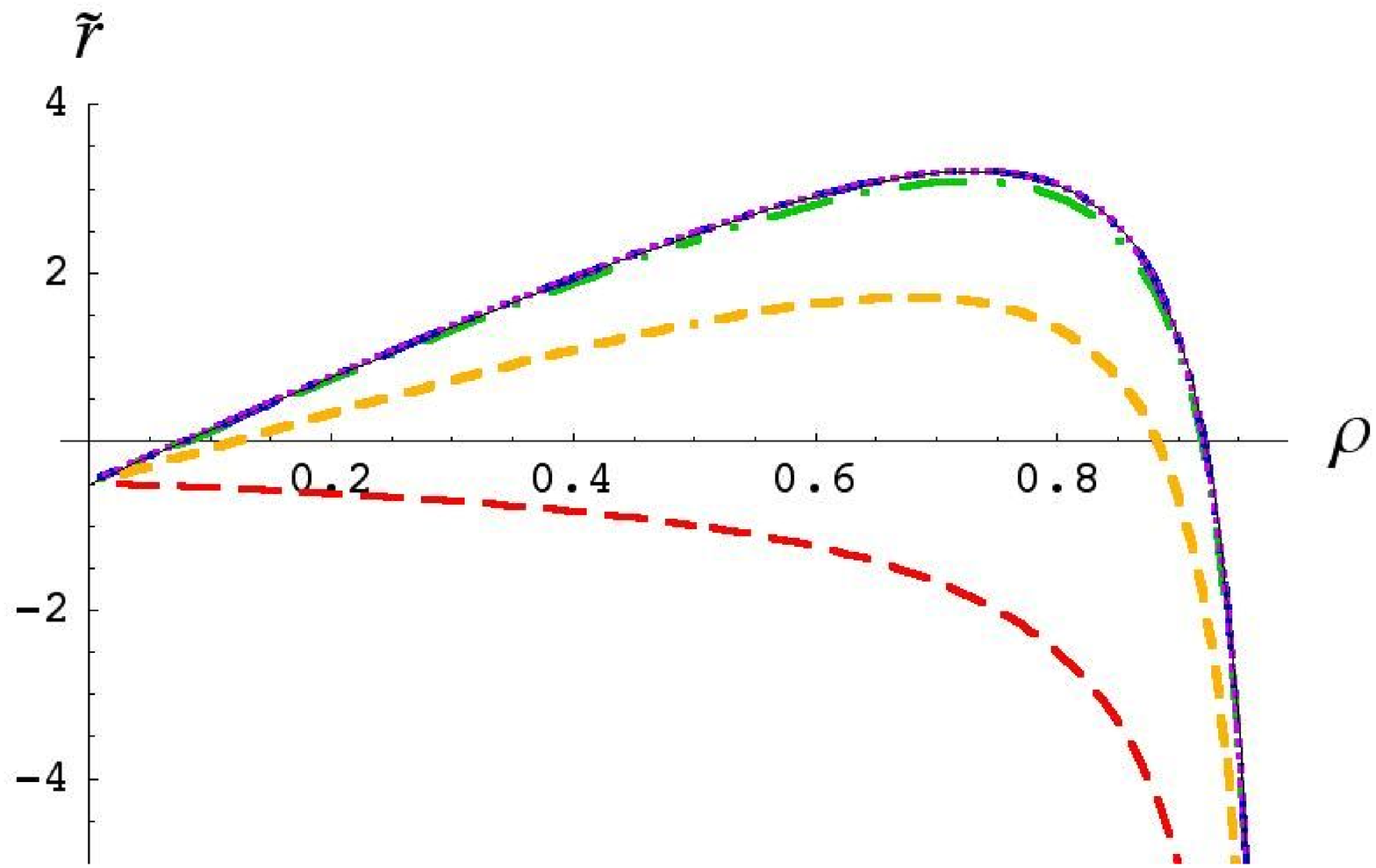}& \\
(e) & (f) & \\
\end{tabular}

\end{center}
\caption{Effects of $a_{max} = (1, 2, 5, 10, 20)$ on $\tilde{r}(\tilde{\rho})$ and $r(\rho)$ (solid black line) for (a) $m_0=0.7$, $T_0=0.45$; (b) $m_0=0.7$, $T_0=0.25$; (c) $m_0=0.9$, $T_0=0.1$; (d) $m_0=0.9$, $T_0=0.01$;  (e)  $m_0=0.01$, $T_0=1$; and (f) $m_0=0.5$, $T_0=0.1$.
}\label{fig:tilde_r_vs_rho} 
\end{figure}

Unlike the baseline model without aging, Eqn. (\ref{eq:EL_allocate1}) does not have an exact closed form analytic solution. Figure \ref{fig:tilde_r_vs_rho} shows a numerical solution for the fitness of the aging bacteria, $\tilde{r}$, as a function of fraction of resources allocated for reproduction, $\tilde{\rho}$, for various values of $a_{max}, m_0$, and $T_0$. The values of $\tilde{r}$ increase drastically between the lowest values of $a_{max}$. However, for higher values of $a_{max}$, $\tilde{r}$ hardly varies and there is a lower return in the fitness for more opportunities to reproduce. Additionally, notice that as $a_{max}$ increases, the peak in $\tilde{r}$ becomes more pronounced, and shifts towards higher values of $\tilde{\rho}$. Thus, bacteria that can double many times gain more fitness from focusing resources for reproduction than those that cannot. 

\begin{figure}[h!]
\begin{center}
\begin{tabular}{c}
\includegraphics[scale=0.225]{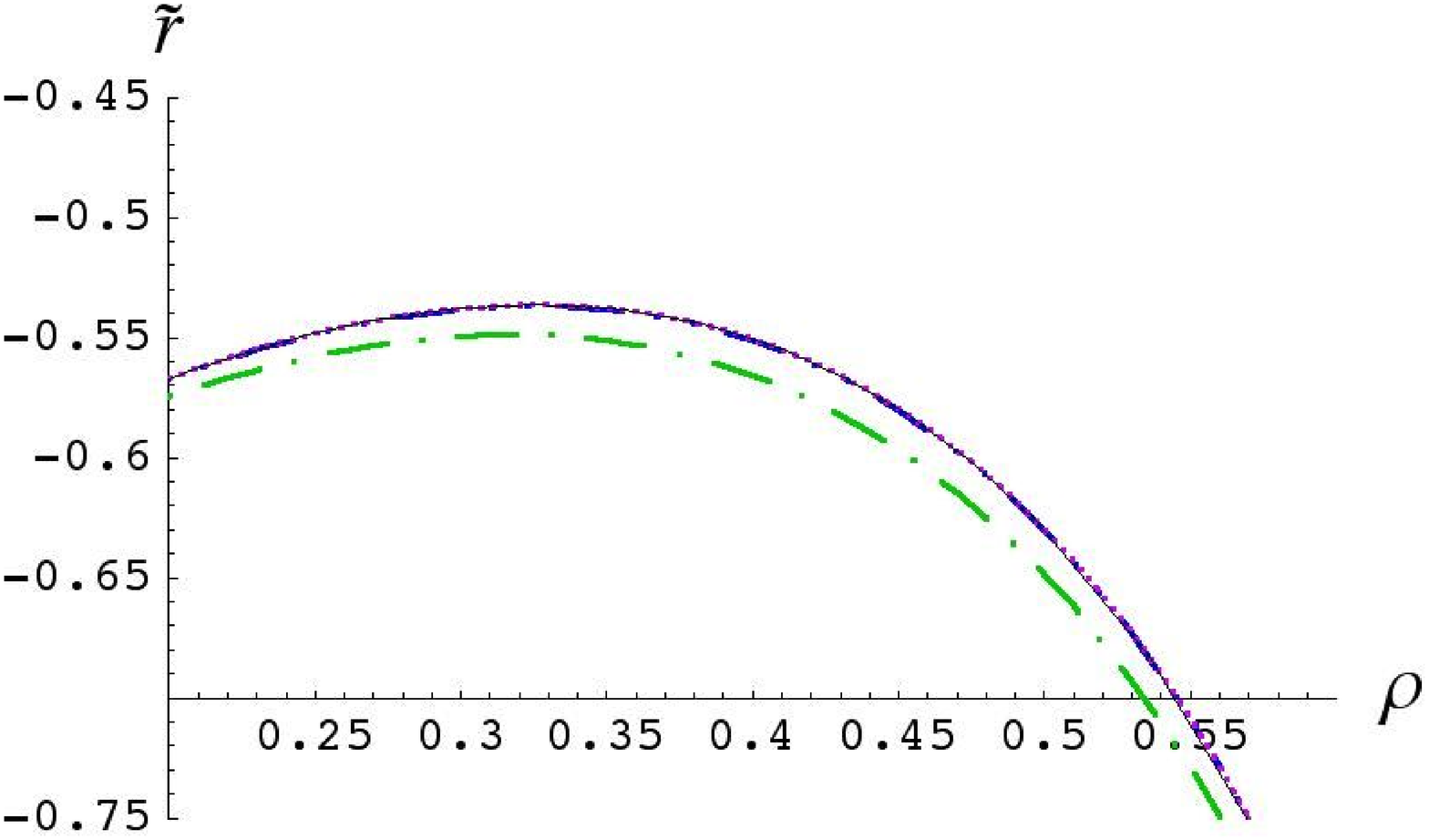} \\
(a) \\
\includegraphics[scale=0.2]{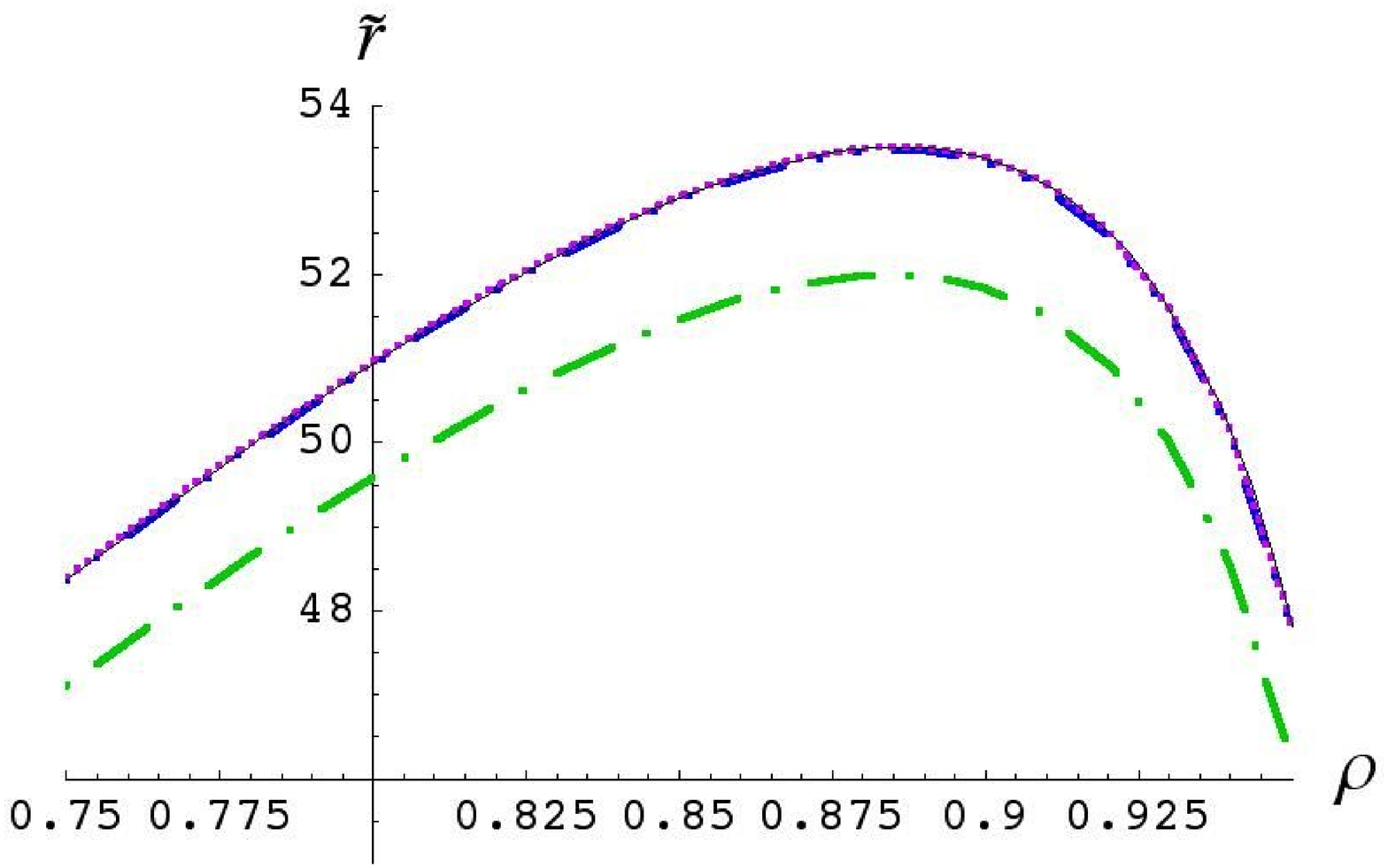} \\
(b) \\
 \\
\includegraphics[scale=0.25]{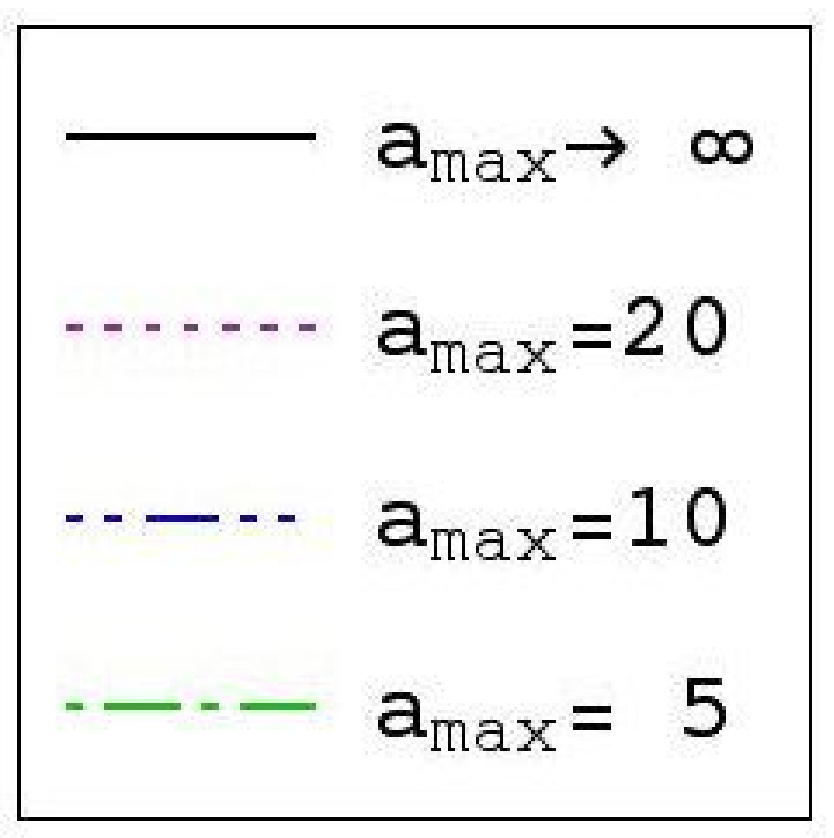}\\
\end{tabular}\\
\end{center}
\caption{Close-ups of the maxima of $\tilde{r}(\tilde{\rho})$ for $a_{max} = (5, 10, 20)$ and $r(\rho)$ (solid black line) for (a) $m_0=0.7$, $T_0=0.25$ (corresponding to Figure \ref{fig:tilde_r_vs_rho}a )  (b) $m_0=0.9$, $T_0=0.01$; (corresponding to Figure \ref{fig:tilde_r_vs_rho}d ) \label{fig:tilde_r_vs_rho_zoom}}
\end{figure}

We can quantify how aging shifts the optimal allocation strategy away from the value determined in Eqn. (\ref{rhostar}) for immortal bacteria. If we differentiate Eqn. (\ref{eq:EL_allocate1}) with respect to $\tilde{\rho}$, and evaluate it at the optimal strategy, $\tilde{\rho}^*$, so that $\frac{d\tilde{r}(\tilde{\rho}^*)}{d\tilde{\rho}} = 0$, we find an implicit expression for the maximum fitness, $\tilde{r}(\tilde{\rho}^*)=\tilde{r}_{max}$, as a function of $\tilde{\rho}^*$:
\begin{equation}
\left( (2 \tilde{\rho}^*-1) m_0 - \tilde{r}_{max}(1-\tilde{\rho}^*)^2 \right) \left[ e^{(\tilde{r}_{max}+ m(\tilde{\rho}^*))T(\tilde{\rho}^*)} +
 a_{max} e^{-(\tilde{r}_{max}+ m(\tilde{\rho}^*))T(\tilde{\rho}^*) a_{max}} \right]  = 0 
 \label{eq:EL_optimal1}
\end{equation}
Since the sum of exponential terms in (\ref{eq:EL_optimal1}) can never be equal to zero, we conclude that
\begin{equation}
(2 \tilde{\rho}^*-1) m_0 - \tilde{r}_{max}(1-\tilde{\rho}^*)^2 = 0, \nonumber
\end{equation}
and therefore $\tilde{r}_{max}$ is related to $\tilde{\rho}^*$ by
\begin{equation}
\tilde{r}_{max} = \frac{(2 \tilde{\rho}^*-1)}{(1-\tilde{\rho}^*)^2} m_0. \label{eq:EL_rmax_implicit}
\end{equation}

It is straightforward to confirm that this expression holds for the case when $a_{max} \rightarrow \infty$. 
As previously discussed, as the maximum cellular age, $a_{max}$, increases, the difference between the fitness of the immortal bacteria, $r$,  and the fitness of the aging bacteria, $\tilde{r}$,  decreases. In particular, for large values of $a_{max}$ we saw that $r \rightarrow \tilde{r}$.  
We can also see in Figure \ref{fig:tilde_r_vs_rho} that when $a_{max} \approx 10$ the peaks in the fitness curves are very close, so that $r_{max} \approx \tilde{r}_{max}$, and occur at about the same value of the allocation strategy, i.e. $\tilde{\rho}^* \approx \rho^*$. We would like to quantify the magnitude of the change in optimal allocation strategy as a function of $a_{max}$. We begin by looking at the difference between $r_{max}$ and $\tilde{r}_{max}$:
\begin{align}
r_{max} - \tilde{r}_{max} & = r(\rho^*) - r(\tilde{\rho}^*) + r(\tilde{\rho}^*) - \tilde{r}(\tilde{\rho}^*). \label{eq:diff_in_r}
 \intertext{Writing (\ref{eq:diff_in_r}), in this way allows us to use our approximation from Eqn. (\ref{eq_Rapprox}) to determine the value of a small parameter, $\epsilon$, in our approximation. Substituting Eqns. (\ref{eq:r_Tm}), (\ref{eq_Rapprox}), and (\ref{eq:EL_rmax_implicit}) into (\ref{eq:diff_in_r}) and rearranging gives:}
 \frac{(2 \tilde{\rho}^*-1)}{(1-\tilde{\rho}^*)^2} m_0 &\approx \frac{\tilde{\rho}^* \ln{2}}{T_0} +\frac{m_0}{1-\tilde{\rho}^*} + \frac{\tilde{\rho}^*}{T_0} \epsilon \nonumber
\end{align}
where $\epsilon = \frac{2^{-a_{max}}}{\ln{2} + a_{max} 2^{-a_{max}}}$. Solving this expression for $\tilde{\rho}^*$, we find:
\begin{equation}
 \tilde{\rho^*} \approx 1-\left(\frac{m_0 T_0}{\ln{2} - \epsilon}\right) ^{\frac{1}{2}} = \rho^* - \frac{1}{2}\left(\frac{\ln{2}}{m_0 T_0}\right)^{\frac{1}{2}} \epsilon + O(\epsilon^2),
\end{equation}
where $O(\epsilon^2)$ denotes terms that are of order $\epsilon^2$ or smaller. From this we also have an approximate expression for $\tilde{r}_{max}$, similar to (\ref{rmax}):
\begin{equation}
\tilde{r}_{max} \approx \frac{1}{T_0} \left( \ln{2} - \epsilon - 2\left(m_0 T_0(\ln{2} - \epsilon)\right)^{\frac{1}{2}} \right). \label{eq:rmax2} 
\end{equation}

This analytic approximation gives results that are very close to those obtained numerically. For instance, for $T_0=0.05$ and $m_0 = 0.5$, 
the difference between the analytic approximation and numerical solution when $a_{max}=5$ is fairly large, as we might have guessed since this is when $r-\tilde{r}$ is largest (Figure \ref{fig:r_tilde_approx}). In this case the numerical solution is $(\tilde{\rho}^*, \tilde{r}_{max}) = (0.8077, 8.31959)$ compared to the analytic approximation $(\tilde{\rho}^*, \tilde{r}_{max}) = (0.804836, 8.00324)$. As $a_{max}$ increases, the approximation improves. For instance, when $a_{max}=20$, the numerical solution is $(\tilde{\rho}^*, \tilde{r}_{max}) = (0.8101, 8.58945)$, and the analytic approximation is $(\tilde{\rho}^*, \tilde{r}_{max}) = (0.810086, 8.59738)$, a difference of less then $0.1\%$ in $\tilde{r}_{max}$ and less than $0.002 \%$ in $\tilde{\rho}^*$.

\section{Discussion}
Studying the impacts of aging and resource allocation on fitness for bacteria is appealing for a number of reasons. Bacterial systems are relatively simple. They also are ideal for experimental manipulation. Metabolism and repair activity should be fairly straightforward to measure, giving indications of how bacteria allocate resources. Genetic manipulation of bacterial systems is also possible, which allows more direct measurement of the parameters in the model. Because generation times are short, bacterial systems also allow for replicate experiments at a reasonable cost. Thus, bacterial systems are ideal model systems for ecology \cite{jessup:2004}.

Here we have proposed a simple mathematical model to explore how aging, in the form of finite life span, effects fitness in bacteria. This allows us to explore, explicitly, predictions of the disposable soma theory of aging for a system that is fairly simple and easy to manipulate. 

This model provides valuable insight into which trade-offs have the most impact on bacterial fitness. Given the assumptions about the form of $l_x$ in the Euler-Lotka equation, as well as the relationships between mortality, doubling rate, and resource allocation, we found that the ability to manipulate the doubling time $T$ has the greatest impact upon bacterial fitness. If the doubling time is short, even if the mortality is high, as is shown in Figure \ref{fig:tilde_r_vs_rho}d, the bacteria's fitness is considerably higher than in a system with lower mortality, but longer doubling time (Figure \ref{fig:tilde_r_vs_rho}e and \ref{fig:tilde_r_vs_rho}f). Thus we expect for there to be strong selection for lower values of the minimum doubling time, $T_0$.

We also find that that a bacteria experiences surprisingly little loss of fitness when it has a finite number of opportunities to reproduce (Figures \ref{fig:tilde_r_vs_amax}  and \ref{fig:tilde_r_vs_rho}). For most combinations of the minimum mortality rate, $m_0$, and minimum doubling time, $T_0$, $a_{max} \approx 5$ or $10$ is large enough to confer almost exactly the same amount of fitness to the bacteria as an ability to reproduce indefinitely (Figure \ref{fig:tilde_r_vs_rho_zoom}). Most of the bacteria's fitness is gained the first few times the it doubles, and the amount gained in each subsequent doubling decreases rapidly. Therefore we expect that if there were a resource cost associated with increasing the maximum number of possible doublings then investing resources to survive to double the first few times would be better than investing additional resources to try to maintain cell integrity indefinitely. 

This simple model also has the advantage of being analytically tractable. Although numerical results for complex models are fairly easy to obtain it can be difficult to understand and quantify the roles of model parameters. Good analytic approximations, such as those found above for $\tilde{r}, \tilde{r}_{max}$, and $\tilde{\rho}^*$, allow us to explore the effects of parameter variation in a very concrete way that might not be available for more complex models. 

This model is only a first step in understanding how aging impacts bacterial fitness. 
Other mechanisms are likely important in determining the optimal life history strategy. The model does not consider the effects of variable environmental conditions, density effects, or of increasing doubling times as a function of age.  Also, here we assume no dependence of $a_{max}$ on $\rho$, so the trade-offs between mortality, reproduction, and lifespan are not explicitly explored. These could be important factors in the development of bacterial life histories. However, in spite of these limitations, this model gives insight into why bacteria do not exhibit infinite lifespans, and suggests directions for further exploration. 

\section{Acknowledgements}
This work was partially supported by a GAANN Fellowship to LJ and NSF Grant DMS 0310542 to MM. 

\nocite{townsend}
\bibliographystyle{apalike}
\bibliography{cholera,biofilm,disease,other,biology}

\end{document}

%% file: lamac.tex
\newcounter{Romancounter}
\newcounter{romancounter}
\newcounter{Alphacounter}
\newcounter{alphacounter}
\newcounter{arabiccounter}